\documentclass[conference]{IEEEtran}

\usepackage{url}
\usepackage{graphicx}
\usepackage{natbib}
\setcitestyle{authoryear,open={(},close={)}}
\usepackage[colorlinks=true,allcolors=black,urlcolor=blue]{hyperref}
\usepackage{cleveref}
\usepackage{adjustbox}
\usepackage{flushend}

\makeatletter
\def\footnoterule{\kern-3\p@
  \hrule \@width \linewidth \kern 2.6\p@} 
\makeatother

\usepackage{tabularx}
\RequirePackage{booktabs}
\usepackage{longtable}
\usepackage{multirow}
\usepackage{tabularx}
\usepackage{float}
\newcolumntype{Y}{>{\centering\arraybackslash}X}
\newcolumntype{C}[1]{@{\hspace{#1}}c@{\hspace{#1}}}

\begin{document}
%
\title{Dynamic Acoustic Unit Augmentation \\With BPE-Dropout for Low-Resource \\End-to-End Speech Recognition}

\author{\IEEEauthorblockN{Aleksandr Laptev\,$^{1,\dagger,*}$, Andrei Andrusenko\,$^{1,\dagger}$, Ivan Podluzhny\,$^{1,\dagger}$, \\Anton Mitrofanov\,$^{1}$, Ivan Medennikov\,$^{1,2,\ddag}$, Yuri Matveev\,$^{1,2,\ddag}$}
\IEEEauthorblockA{$^{1}$Corporate Laboratory of Human-Machine Interaction Technologies, \\Information Technologies and Programming Faculty, \\School of Translational Information Technologies, \\ITMO University, Saint-Petersburg, Russia 197101}
\IEEEauthorblockA{$^{2}$STC-innovations Ltd, Saint-Petersburg, Russia 194044}
\IEEEauthorblockA{$^{*}$Corresponding author: Aleksandr Laptev, \href{mailto:aalaptev@itmo.ru}{aalaptev@itmo.ru}}}

\maketitle

\begin{abstract}
With the rapid development of speech assistants, adapting server-intended automatic speech recognition (ASR) solutions to a direct device has become crucial. Researchers and industry prefer to use end-to-end ASR systems for on-device speech recognition tasks. This is because end-to-end systems can be made resource-efficient while maintaining a higher quality compared to hybrid systems. However, building end-to-end models requires a significant amount of speech data. Another challenging task associated with speech assistants is personalization, which mainly lies in handling out-of-vocabulary (OOV) words. In this work, we consider building an effective end-to-end ASR system in low-resource setups with a high OOV rate, embodied in Babel Turkish and Babel Georgian tasks. To address the aforementioned problems, we propose a method of dynamic acoustic unit augmentation based on the BPE-dropout technique. It non-deterministically tokenizes utterances to extend the token's contexts and to regularize their distribution for the model's recognition of unseen words. It also reduces the need for optimal subword vocabulary size search. The technique provides a steady improvement in regular and personalized (OOV-oriented) speech recognition tasks (at least 6\% relative WER and 25\% relative F-score) at no additional computational cost. Owing to the use of BPE-dropout, our monolingual Turkish Conformer established a competitive result with 22.2\% character error rate (CER) and 38.9\% word error rate (WER), which is close to the best published multilingual system.
\end{abstract}


\begingroup\renewcommand\thefootnote{$\dagger$}
\begin{NoHyper}
\footnotetext{These authors have contributed equally to this work and share first authorship}
\end{NoHyper}
\endgroup

\begingroup\renewcommand\thefootnote{$\ddag$}
\begin{NoHyper}
\footnotetext{These authors share senior authorship}
\end{NoHyper}
\endgroup

\section{Introduction}
Digital speech assistants are becoming ubiquitous in everyday life. According to the survey from Microsoft’s latest voice report \citep{olson2019voice}, 75\% of English-speaking households are expected to have at least one smart speaker by 2020. One of the key functions of an ordinary speech assistant is voice search. It allows users to search the web by saying queries rather than typing them. Besides, voice search is expected to be as personalized as any modern search. However, the personalization itself is more complicated for this task than for typing-based search since it starts before ranking the results at the speech recognition stage. The part of voice search that is responsible for transducing speech to words and passing them to the search field can be thought of as the large vocabulary continuous speech recognition (LVCSR) task of automatic speech recognition (ASR). One of the main problems in this task is the recognition of words that the ASR system has not encountered before, called out-of-vocabulary (OOV) words. Recognition errors for such words occur significantly more often than those that the system is aware of. Thus, the presence of OOV words in voice queries may negatively affect the performance of voice search. In turn, an incorrect voice search may decrease the user-perceived quality of the whole system. Moreover, in general, speech assistants' low personalization ability often leads to deterioration of user experience \citep{pal2019user}.

ASR system is one of the main components of a smart voice assistant. This system recognizes speech information from the user in order to transform it and pass it on for processing as a command or query. Thus, recognition errors can lead to incorrect interpretation of commands or incorrect formation of search queries. But to operate effectively, it is not enough for the ASR model to have a high recognition quality. It also has to be fast and compact to be able to run on edge devices \citep{sainath2020streaming,huang2020} or to have a combined server-device structure with a lightweight model for commands and a high-quality server-grade LVCSR-intended model \citep{sigtia2018}. To our knowledge, both hybrid \citep{hinton_hmm-dnn} and end-to-end \citep{graves_connectionist_2006,chan2015listen} ASR systems are used for speech assistants \citep[e.g.][]{aleksic2015improved,tulsiani2020}. But regardless of technology, building an ASR system for smart assistants faces the data availability problem. Its essence is, due to speech data privacy concerns and the existence of underrepresented languages, there might be challenging to gather enough data to build an effective recognition system. Thus, for many languages, excluding English, one has to consider low-resource data availability conditions (the total amount of data is less than a hundred hours).

The aforementioned problems of OOV words handling and low-resource data conditions need to be addressed when building an ASR system. If the system is a conventional hybrid (HMM-DNN-based acoustic model and word-based n-gram language model), the OOV problem is often solved by dynamic expanding the system's vocabulary and/or adapting the language model \citep[e.g.][]{Khokhlov2017,gandhe18,malkovsky2020techniques}. A less common approach is to use a subword-based n-gram language model \citep{Smit2017}. The vocabulary of character- or subword-based end-to-end systems is not restricted, as it is for the hybrid. However, it is challenging to build a model using extra unpaired data (viz. large external text corpora), and doing this can lead to poor performance on rare and unseen words. One of the recent approaches to tackle the OOV problem for such systems is biasing towards a given context at decoding time \citep{jain2020contextual}. However, even without such improvements, subword-based end-to-end systems are generally better in handling OOV than conventional hybrid ones. The downside is that the negative impact of low-resource conditions affects end-to-end systems more since the acoustic units (output tokens) of such systems are more high-level than the Hidden Markov Model states of hybrid ones. In other words, there are more data required to saturate the model (without noticeable overfitting) that emits end-to-end acoustic units. Concerning the choice of acoustic units for an end-to-end ASR system, there is a tradeoff between better saturation, obtained through the use of less specific tokens, and higher token precision by using more specific and curated tokens that are expected to contain non-trivial lexical information. Excluding from consideration logogram-based languages (e.g., Chinese), characters are the least specific tokens, and various word pieces (subwords) are more specific ones.

There are many ways to divide words into subwords. The two most popular methods of subword segmentation are Byte Pair Encoding (BPE) \citep{sennrich-etal-2016-neural} and a unigram subword language model \citep{kudo-2018-subword}. BPE is a process of agglomerative merging of subwords, starting with characters, according to the frequency of their joint occurrence in the training set. Unigram language model (ULM) subword segmentation is an approach for inferring subword units by training a unigram language model on a set of characters and words suffix arrays and iteratively filtering out subwords using the Expectation-Maximization algorithm to maximize the data likelihood. But this approach to make the ULM subword segmentation is not the only one. Another method worth mentioning is Morfessor \citep{creutz-lagus-2002-unsupervised}, which finds morphological segmentation of words using greedy local search. Regardless of the subword segmentation method, there is a problem of finding the optimal (in terms of the final system performance) subword number. Another problem related to subword usage is the variability of their segmentation. A text segmented with the smallest number of highly specific subwords may not always be optimal. We propose using dynamic acoustic unit augmentation to address these problems. The approach consists of diversifying the subword segmentation during model training by sampling different segmentations for the same words. Such sampling is supported in the ULM by simply varying the sampling temperature. It is called the subword regularization \citep{kudo-2018-subword}. A recent modification of Morfessor, named Morfessor EM+Prune \citep{gronroos2020morfessor} is also able to perform the subword regularization. Finally, BPE-dropout \citep{provilkov-etal-2020-bpe} was proposed to regularize segmentation by randomly omitting merges.

There are few previous works on ASR related to the investigation of subword augmentation by non-deterministic segmentation. The vanilla subword regularization was studied in \citep{drexler2019} and \citep{Lakomkin2020}. In the first work, the method was applied for the WSJ dataset (English, 50 hours). Also, the authors proposed a novel prefix search algorithm that utilizes subword length in the calculation of prefix probability. The second work investigated the improvement of applying the subword regularization to different amounts of data and analyzed its effect on OOV word recognition and hypothesis diversity. Presently, BPE-dropout and Morfessor EM+Prune were applied only to machine translation (MT). BPE-dropout was beneficially used for low-resource MT tasks as a standalone improvement \citep{tapo2020neural,knowles-etal-2020-nrc-systems,libovicky-etal-2020-lmu} or combined with a neural sequence-to-sequence segmentation model \citep{he-etal-2020-dynamic}. The Morfessor EM+Prune's subword regularization, along with other improvements, was used for asymmetric-resource one-to-many MT task \citep{gronroos2020transfer}.

In this work, we provided extensive research on how BPE-dropout and ULM subword regularization acoustic unit augmentations contribute to the performance of strong end-to-end ASR system baselines in low-resource conditions. We studied the sensitivity of a model to the total number of target subwords and the regularization rate. We also analyzed how effective the aforementioned subword augmentation techniques are for alleviating the OOV problem.

Our contribution is three-fold:
\begin{itemize}
  \item We proposed and evaluated a dynamic acoustic unit augmentation method for ASR system training, making speech assistants' user experience and perceived quality better by improving the OOV word recognition quality. The method is based on a non-deterministic BPE subword segmentation algorithm, named BPE-dropout.
  \item We compared and analyzed WER reduction and OOV handling ability of BPE-dropout with those of ULM subword regularization.
  \item Finally, we built systems that achieved competitive results for IARPA Babel Turkish \citep{babel-turkush-dataset} and Georgian \citep{babel-georgian-dataset} low-resource tasks.
\end{itemize}
\section{ASR modeling}
\label{sec:asr}
This section provides an overview of two main ASR approaches: hybrid and end-to-end.

\subsection{Hybrid approach}

Conventional hybrid ASR systems can be divided into acoustic and language models. The acoustic model is responsible for converting input feature sequence to output acoustic units (e.g., phonemes). The language model contains the language's knowledge and helps the decoder convert acoustic units into the final word sequence. Apart from a few service parts, it includes pronunciation lexicon and linguistic information, applied as a statistical n-gram model. The pronunciation lexicon defines the rules for mapping graphemes (characters) to phonemes.

In recent years, hybrid systems have been well studied and proven to solve many ASR-related problems. However, this approach to training ASR systems has inherent drawbacks:
\begin{itemize}
    \item Acoustic and language models are built separately from each other and have their different objective functions. This significantly complicates the process of optimizing the ASR system.
    \item To train the final DNN-based acoustic model, hard alignment (mapping of each input feature frame to a target acoustic unit) is required. It is generated and refined through several iterations of GMM-HMM-based training, in which the condition-independent assumption is in effect. But this hard alignment also limits the acoustic context that the model can process before emitting the target token's spike.
    \item Decoding with WFST graph. This process is highly memory intensive, making it difficult to use in ASR tasks for smart devices where the memory is severely limited.
\end{itemize}

\subsection{End-to-end approaches}

\textbf{CTC}. Connectionist Temporal Classification (CTC) \citep{graves_connectionist_2006} was the first significant step towards addressing hybrid models' problems mentioned earlier. A new loss function that allows mapping of input features to final speech recognition labels without using hard alignment and pronunciation lexicon was proposed. Any acoustic units (graphemes, phonemes, subwords) can be used as output labels. An auxiliary “blank” symbol controls label repetitions and their absence. However, the CTC-trained end-to-end ASR system does not have its own context- or language model (the system is encoder only, and it is trained in a context-independent manner), which leads to degradation of the recognition quality. Nevertheless, using pure CTC-trained systems can still be advantageous since such systems are often the most efficient and deliver competitive quality \citep{kriman2020quartznet}.

\textbf{Neural transducer}. Later, the neural transducer \citep{graves2012transducer} was introduced, which can solve the context-independent problem of the CTC approach. The proposed prediction network is designed to utilize contextual information and thus works similarly to the language model. The encoder and prediction network results are then sent to the joint network, which emits the final result based on acoustic and context information. The entire system is jointly optimized with a single Transducer objective function, which is a modified CTC loss. Recently, the transducer approach has proven its effectiveness both in large-resource \citep[e.g.][]{li2020developing} and low-resource \citep[e.g.][]{andrusenko2020towards} tasks.

\textbf{Attention-based}. Another approach to building an end-to-end ASR system is using the attention-based sequence-to-sequence architecture \citep{chan2015listen} that consists of an encoder and a decoder with the attention mechanism. The attention mechanism allows the decoder to use a weighted representation of encoded input context. This, along with an autoregressive decoder, provides context-depending label modeling. However, this approach is prone to overfitting, which manifests in the output of a highly probable sequence of tokens regardless of acoustic information. It was proven effective to combine attention decoding with CTC to alleviate their shortcomings and improve recognition quality \cite{kim_joint_2017}. At the same time, the Transformer \citep{attention_2017} attention-based architecture was proposed as more effective than various RNN architectures. A multi-head self-attention (MHA) mechanism significantly improved the quality of models over the recurrent models. A transformer model, trained with CTC-Attention, can outperform neural transducer systems \citep[e.g.][]{andrusenko2020exploration} and benefit from various augmentation techniques \citep{laptev2020need}. Recently, The Conformer \citep{Gulati2020} was introduced, which is a modification of the transformer layer. Convolution blocks and advanced activation functions were added to each layer of the model encoder. Latest reports \citep[e.g.][]{guo2020recent} demonstrate that the Conformer outperforms Transformer in almost all tasks.

\section{Method description}
\label{sec:method}
This section describes the subword augmentation techniques that were the subject of our investigation. An evaluation method for the recognition performance of OOV words is also provided here.

\subsection{ULM subword regularization}
The subword segmentation algorithm \citep{kudo-2018-subword} is based on a simple unigram language model. It allows getting multiple subword segmentation variants with the corresponding probabilities. The probability of a subword sequence \(\mathbf{x} = (x_1, x_2, ..., x_M)\) is the product of unigram probabilities of these subwords. To obtain the most probable subword segmentation \(\mathbf{x^{*}}\) for the input word sequence \(\mathbf{W}\) the Viterbi algorithm is used.

For subword regularization, one first has to get \textit{l}-best segmentations according to a probability distribution \(P(\mathbf{x}|\mathbf{W})\) over subword segmentation variants corresponding to a source word sequence. Next, one can sample a new segmentation \(\mathbf{x}_i\) from the multinomial distribution:
\begin{equation}
P(\mathbf{x}_i|\mathbf{W}) = \frac{P(\mathbf{x}_i)^\alpha}{\sum_{i=1}^{l} P(\mathbf{x}_i)^\alpha}
\end{equation}
where \(\alpha\) is a temperature parameter, which controls the smoothness of the distribution. In case of \(\alpha = 0\), the segmentation is sampled from uniform distribution (segmentation is uniformly sampled from the \textit{n}-best (if $l=n$) or lattice (if $l=\infty$)). A larger \(\alpha\) allows the selection of the most probable Viterbi segmentation. The parameter \textit{l} is restricted by Forward-Filtering and Backward-Sampling algorithm \citep{scott2002a} because the number of all possible subword segmentation variants increases exponentially with respect to the sentence length.

\subsection{BPE-dropout augmentation}
The Byte Pair Encoding (BPE) \citep{sennrich-etal-2016-neural} segmentation defines a simple deterministic mapping of words to subword tokens. The algorithm starts by creating an initial token vocabulary consisting of characters of the input text's words. The end-of-word mark is also added to disambiguate word boundaries. Next, the tokens are agglomeratively merged according to their co-occurrence frequency. The merge operations are written in the merge table. The algorithm iterates until the maximum number of merges is exceeded, or the desired vocabulary size is reached. The resulting merge table contains all allowed rules and order of merging subwords.

During the segmentation process, the word is split into characters with the addition of the end-of-word mark. Then the tokens are assembled according to the merge table until the merge rules are exhausted. The training and inference procedures are deterministic, thus the result of the algorithm is always unambiguous. Such formulation does not imply any regularization or augmentation.

Recently, BPE was reformulated \citep{provilkov-etal-2020-bpe}, which made applying augmentation possible. The method, named BPE-Dropout, is based on randomly discarding a certain number of merges with some probability \(p\). If \(p = 0\), then it operates like standard BPE segmentation. In the case of \(p = 1\), all merge operations are omitted, and words are split into single characters.

\subsection{Recognition of OOV words}
It is assumed that an end-to-end ASR model trained on subword-segmented utterances is capable to recognize any new word in the target language. However, if a word was not sufficiently represented in the training data, then the model might assign a low probability to the subword sequence representing the word during the decoding process. Therefore, instead of an OOV word, the decoder is likely to emit the most similar seen word. We assumed that non-deterministic subword tokenization should improve the recognition of unseen words, as this technique allows enhancing the diversity of subword sequences during the ASR model's training process.

To analyze OOV word recognition performance, we used an F-score metric similar to \citep{Lakomkin2020}. The method based on counting after decoding how many times the model emitted (true positive, \(tp\)) or did not emit (false negative, \(fn\)) the OOV words from the evaluation set. Words that were neither in training nor in evaluation transcripts (false positive, \(fp\)) are also counted and used for calculating \(precision = tp / (tp + fp)\) , \(recall = tp / (tp + fn)\), and \(\textit{F-score} = 2\cdot  precision\cdot recall/(precision+recall)\), which allows to estimate the quality of OOV word recognition.

\section{Experiments}
\label{sec:exp}
This section describes the experiments performed and provides the obtained results.

\subsection{Data}
For our experiments, we used two telephone conversations datasets for IARPA Babel Turkish \citep{babel-turkush-dataset} and Georgian \citep{babel-georgian-dataset} languages. We formed training sets from utterances with a duration from 10 to 2000 frames and not exceeding 300 characters to avoid GPU memory overflow and stabilize the training process. We also extracted one hour of data from each training set for validation purposes. Final data training sizes were 73.40 hours for Turkish and 50.52 hours for Georgian sets. All results were made for the official development sets, which consist of 9.82 hours (5.40\% OOV words) and 12.36 hours (8.95\% OOV words) of Turkish and Georgian, respectively.

\subsection{End-to-end setup}
The main end-to-end model architecture for our experiments was Transformer. The encoder consisted of a 2-layer conv2d subsampling block (to reduce input feature sequence by four times) followed by 12 Transformer layers with 1024 units feed-forward dimension. The decoder was a 6-layer Transformer with 1024 feed-forward units. We used 8-headed self-attention with 360 dimensions for both model parts. The model was trained with joint CTC and attention-based loss for 100 epochs. We used OneCycleLr training scheduler \citep{smith2019super} as it showed the best model convergence during a preliminary architecture search. The input feature sequence for both hybrid and end-to-end setups were cepstral mean and variance normalized 40-dimensional log-Mel filterbank coefficients with 3-dimensional pitch features. In our end-to-end training setup, we additionally used SpecAugment \citep{Park_2019} data augmentation.

To the extent of our knowledge, the only tool that currently supports both the ULM and BPE subword segmentation algorithms and their non-deterministic segmentation techniques is the Sentencepiece tokenizer \citep{kudo-richardson-2018-sentencepiece}. We used it to dynamically tokenize utterances when training our models in the ESPnet speech recognition toolkit \citep{Watanabe2018,watanabe2020}.

\subsection{The augmentation impact}
\begin{figure*}[ht]
  \centering
  \includegraphics[scale=1]{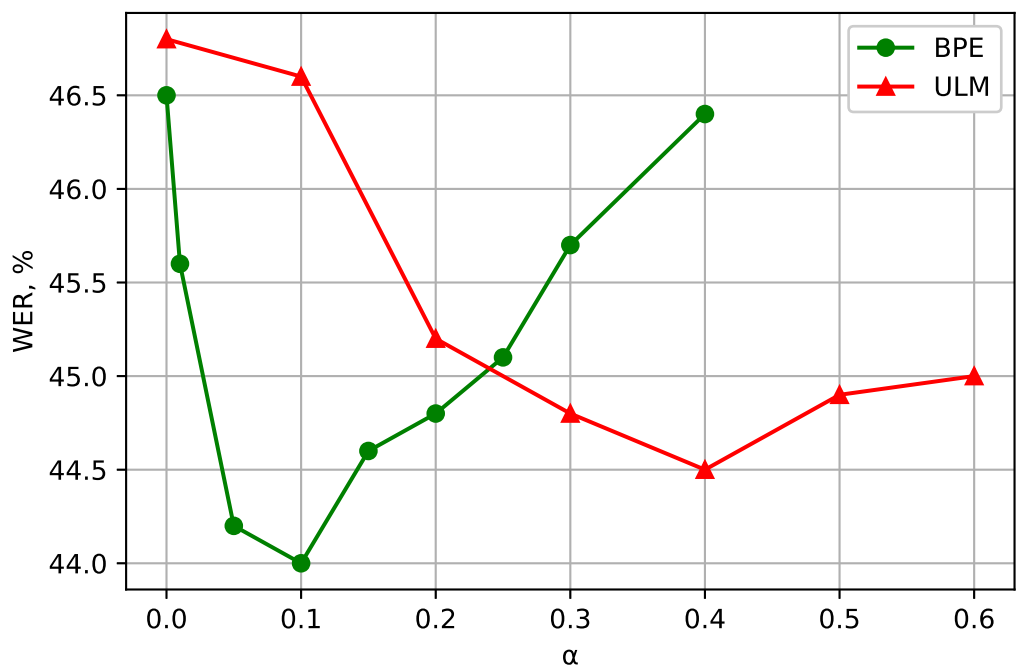}
  \caption{(Turkish) WER for different values of \(\alpha\) for BPE and ULM subword models. \(\alpha = 0.0\) means that deterministic segmentation is used. The vocabulary size is 1000 units.}
  \label{fig:1}
\end{figure*}

\Cref{fig:1} shows how Word Error Rate (WER) depends on the usage of augmentation techniques. In the first series of experiments, for the Turkish language, we trained the Transformer model described above using two different subword tokenization methods: BPE and ULM. The vocabulary size was set to 1000 units. For each method, a line graph of the dependence of \(\alpha\) value (ULM sampling smoothing parameter and BPE dropout probability of a subword segmentation model in the Sentencepiece tokenizer) on WER (green and red colors respectively) was plotted. The scale mark \(\alpha = 0\) denotes deterministic tokenization. It can be observed that using both augmentation methods was beneficial for models. The best result was obtained using the BPE model with \(\alpha = 0.1\), which provided an absolute WER improvement of 2.5\%.

\begin{table}[ht]
  \centering
  \caption{(Turkish): Different vocabulary sizes}
  \label{tab:vocab_size_turkish}
  \begin{adjustbox}{width=1\linewidth}
  \begin{tabular}{ c c c c c c }
    \toprule
    \textbf{Vocab size} & \textbf{BPE-dropout} & \textbf{WER, \%} & \textbf{Precision} & \textbf{Recall} & \textbf{F-score} \\
    \midrule
    char & - & 53.0 & 0.067 & 0.165 & 0.095 \\
    \midrule
    \multirow{2}{*}{500} & - & 46.1 & 0.114 & 0.152 & 0.130 \\
    & + & 44.4 & 0.120 & 0.209 & 0.153 \\
    \midrule
    \multirow{2}{*}{1000} & - & 46.5 & 0.130 & 0.144 & 0.137 \\
    & + & 44.0 & 0.126 & 0.194 & 0.153 \\
    \midrule
    \multirow{2}{*}{2000} & - & 47.4 & 0.126 & 0.118 & 0.123 \\
    & + & 43.6 & 0.144 & 0.198 & 0.167 \\
    \midrule
    \multirow{2}{*}{3000} & - & 49.2 & 0.129 & 0.099 & 0.112 \\
    & + & 43.2 & 0.156 & 0.197 & 0.174 \\
    \midrule
    \multirow{2}{*}{4000} & - & 50.9 & 0.124 & 0.085 & 0.101 \\
    & + & 43.4 & 0.151 & 0.183 & 0.166 \\
    \midrule
    \multirow{2}{*}{5000} & - & 51.1 & 0.115 & 0.070 & 0.087 \\
    & + & 45.0 & 0.137 & 0.160 & 0.148 \\
    \bottomrule
  \end{tabular}
  \end{adjustbox}
\end{table}

\begin{table}[ht]
  \centering
  \caption{(Georgian): Different vocabulary sizes}
  \label{tab:vocab_size_georgian}
  \begin{adjustbox}{width=1\linewidth}
  \begin{tabular}{ c c c c c c }
    \toprule
    \textbf{Vocab size} & \textbf{BPE-dropout} & \textbf{WER, \%} & \textbf{Precision} & \textbf{Recall} & \textbf{F-score} \\
    \midrule
    char & - & 51.2 & 0.090 & 0.162 & 0.116 \\
    \midrule
    \multirow{2}{*}{100} & - & 50.2 & 0.087 & 0.143 & 0.108 \\
    & + & 48.5 & 0.101 & 0.167 & 0.126 \\
    \midrule
    \multirow{2}{*}{500} & - & 49.5 & 0.095 & 0.126 & 0.108 \\
    & + & 46.3 & 0.117 & 0.172 & 0.140 \\
    \midrule
    \multirow{2}{*}{1000} & - & 50.7 & 0.088 & 0.096 & 0.092 \\
    & + & 46.6 & 0.118 & 0.161 & 0.137 \\
    \midrule
    \multirow{2}{*}{2000} & - & 53.8 & 0.070 & 0.061 & 0.066 \\
    & + & 47.1 & 0.124 & 0.160 & 0.140 \\
    \midrule
    \multirow{2}{*}{3000} & - & 57.0 & 0.054 & 0.039 & 0.046 \\
    & + & 47.9 & 0.116 & 0.147 & 0.130 \\
    \bottomrule
  \end{tabular}
  \end{adjustbox}
\end{table}

Having settled on the BPE-dropout with \(\alpha = 0.1\) dropout probability, we investigated how this augmentation technique performs for different vocabulary sizes. The results for the Turkish and Georgian languages are presented in \Cref{fig:2,fig:3} respectively (a more detailed representations are available in \Cref{tab:vocab_size_turkish,tab:vocab_size_georgian}). Models with character-based acoustic units performed worse (with 53.0 and 51.2 WER\% for Turkish and Georgian) than ones with subword-based optimal vocabulary sizes. Despite this, character-based models had a high recall and thus a competitive OOV recognition F-score. Using the chosen unit augmentation technique was beneficial both in WER and F-score. With the vocabulary size of 3000, Turkish recognition quality was improved by 2.9 WER\% and 0.034 F-score compared to the best non-augmented models and by 6.0 WER\% and 0.062 F-score for those of the same vocabulary size. Similarly, improvements of 3.2 WER\% and 0.032 F-score were obtained for the Georgian language with 500 acoustic units. Overall, the use of BPE-dropout relaxed the need for optimal subword vocabulary size choice to build a more effective model.

Another study was to check the BPE-dropout augmentation when applied with a more advanced Conformer architecture and other augmentation approaches. We chose a Conformer with the depth-wise convolution kernel of size 15 and the 3-fold speed perturbation (SP). The tokenization setup was as follows: 3000 BPE vocabulary units and the dropout probability \(\alpha = 0.1\). The results for Babel Turkish are presented in \Cref{tab:transformer_conformer}. The BPE-dropout augmentation improvement remained for the Conformer setup with 2.4 WER\% and 0.035 F-score improvement compared to 6.0 WER\% and 0.064 F-score of the Transformer improvement. It was also productively combined with the SP augmentation, resulting in 38.9 WER\% and 0.224 F-score of the final system. The training of the Conformer model did not converge for Babel Georgian in our setup. It can be assumed that 50 hours of data may not be enough to train an advanced end-to-end model from scratch.

\begin{table}[h!]
  \centering
  \caption{(Turkish): end-to-end model performance}
  \label{tab:transformer_conformer}
  \begin{adjustbox}{width=1\linewidth}
  \begin{tabular}{c | c | c | c c c}
    \toprule
    \textbf{model} & \textbf{BPE-dropout} & \textbf{WER, \%} & \textbf{Precision} & \textbf{Recall} & \textbf{F-score} \\
    \midrule
    \multirow{2}{*}{Transformer} & - & 49.2 & 0.129 & 0.099 & 0.112 \\
    & + & 43.2 & 0.156 & 0.197 & 0.174 \\
    \midrule
    \multirow{2}{*}{Conformer} & - & 42.9 & 0.188 & 0.142 & 0.162 \\
    & + & 40.5 & 0.194 & 0.201 & 0.197\\
    \midrule
    Conformer+SP & + & 38.9 & 0.199 & 0.255 & 0.224 \\
    \bottomrule
  \end{tabular}
  \end{adjustbox}
\end{table}

\begin{table*}[h!]
  \centering
  \caption{Final comparison. * indicates that sclite is used for scoring}
  \label{tab:final_comparison}
  \begin{adjustbox}{width=0.8\linewidth}
  \begin{tabularx}{0.7\linewidth}{ Y | C{15pt} | Y  Y }
    \toprule
    \textbf{Language} & \textbf{Model} & \textbf{CER, \%} & \textbf{WER, \%} \\
    \midrule
    \multirow{6}{*}{Turkish} & Our LF-MMI TDNN-F & (*21.4) & 43.9 (*38.6) \\
    & Our Conformer & 22.2 (*17.3) & 38.9 (*34.7) \\
    & CTC-BLSTM \citep{bataev_2018} & - & 50.7 (*45.8) \\
    & BLSTMP+VGG-Multilingual \citep{Cho2018Multilingual} & 28.7 & - \\
    & XLSR-Monolingual \citep{conneau2020unsupervised} & 26.1 & - \\
    & XLSR-53-Multilingual \citep{conneau2020unsupervised} & 18.8 & - \\
    \midrule
    \multirow{6}{*}{Georgian} & Our LF-MMI TDNN-F & (*25.4) & 51.6 (*43.3) \\
    & Our Transformer & 24.6 (*21.0) & 46.3 (*41.7) \\
    & BLSTMP+VGG-Multilingual \citep{Cho2018Multilingual} & 36.0 & - \\
    & Multilingual hybrid fusion \citep{Alum2017Georgian} & - & 32.2 \\
    & XLSR-Monolingual \citep{conneau2020unsupervised} & 30.5 & - \\
    & XLSR-53-Multilingual \citep{conneau2020unsupervised} & 17.2 & 31.1 \\    
    \bottomrule
  \end{tabularx}
  \end{adjustbox}
\end{table*}

\subsection{Final comparison}
Apart from our best end-to-end systems, we established baselines with a conventional hybrid architecture consisting of an LF-MMI trained TDNN-F acoustic model and a 3-gram word language model. The acoustic features were the same we used for our end-to-end models. The models setup and training process (except for acoustic features) were performed according to the \textit{librispeech/s5} recipe of the Kaldi \citep{Povey11thekaldi} toolkit.

Our baselines and best models for both languages were compared with other published results in \Cref{tab:final_comparison}. There is a specific type of recognition result scoring, named sclite\footnote{sclite is a part of the SCTK toolkit. Available at \url{https://github.com/usnistgov/SCTK}} \citep{fisher1993alignment}. It was used in all NIST OpenKWS evaluations and provided for all BABEL languages. Thus, the considered Babel Turkish and Georgian development sets are expected to be scored with it. But since all works except \citep{bataev_2018} do not mention the use or non-use of the sclite scoring tool, the exact comparison is not formally possible. Results that are known to have been sclite-scored are marked with an asterisk.

Our Turkish end-to-end model performed well across all systems. It delivered 22.2\% character error rate (CER) and 38.9\% WER (17.3\% CER and 34.7\% WER with sclite scoring). These results are better than those of previous monolingual systems. The model may even have outperformed the best Babel multilingual system (assuming sclite was used in \citep{conneau2020unsupervised}) in CER. This might indicate that applying advanced data augmentation techniques can compete with out-of-language-domain data addition in terms of quality improvement. However, there are currently few works covering Turkish speech recognition, so the topic has yet to be fully explored. As for Babel Georgian, our model with 24.6\% CER and 46.3\% WER (21.0\% CER and 41.7\% WER with sclite scoring) was competitive among monolingual setups, but considerably below previous multilingual results. Apart from out-of-language-domain data usage, this gap can be explained by additional text data usage in building language model for decoding \citep{Alum2017Georgian} and advanced multilingual pre-training approaches \citep{conneau2020unsupervised}.

\section{Discussion}
\label{sec:dis}
This section attempts to explain the results provided in Experiments (\Cref{sec:exp}).

With a subword text segmentation, tokens can be unevenly represented in the training data, and a model can be biased towards recognizing frequent tokens. Nevertheless, even frequent tokens can have a small limited number of words of which they are a part (context words), and this can lead to overfitting. Also, short (in terms of character number) tokens in such conditions may have poor saturation, especially in low-resource cases. BPE-dropout can address both of these problems: it increases the frequency of short tokens and the number of context words for all tokens (except for subwords representing the full word) in the training process.

\begin{figure*}[h!]
  \centering
  \includegraphics[width=1\textwidth]{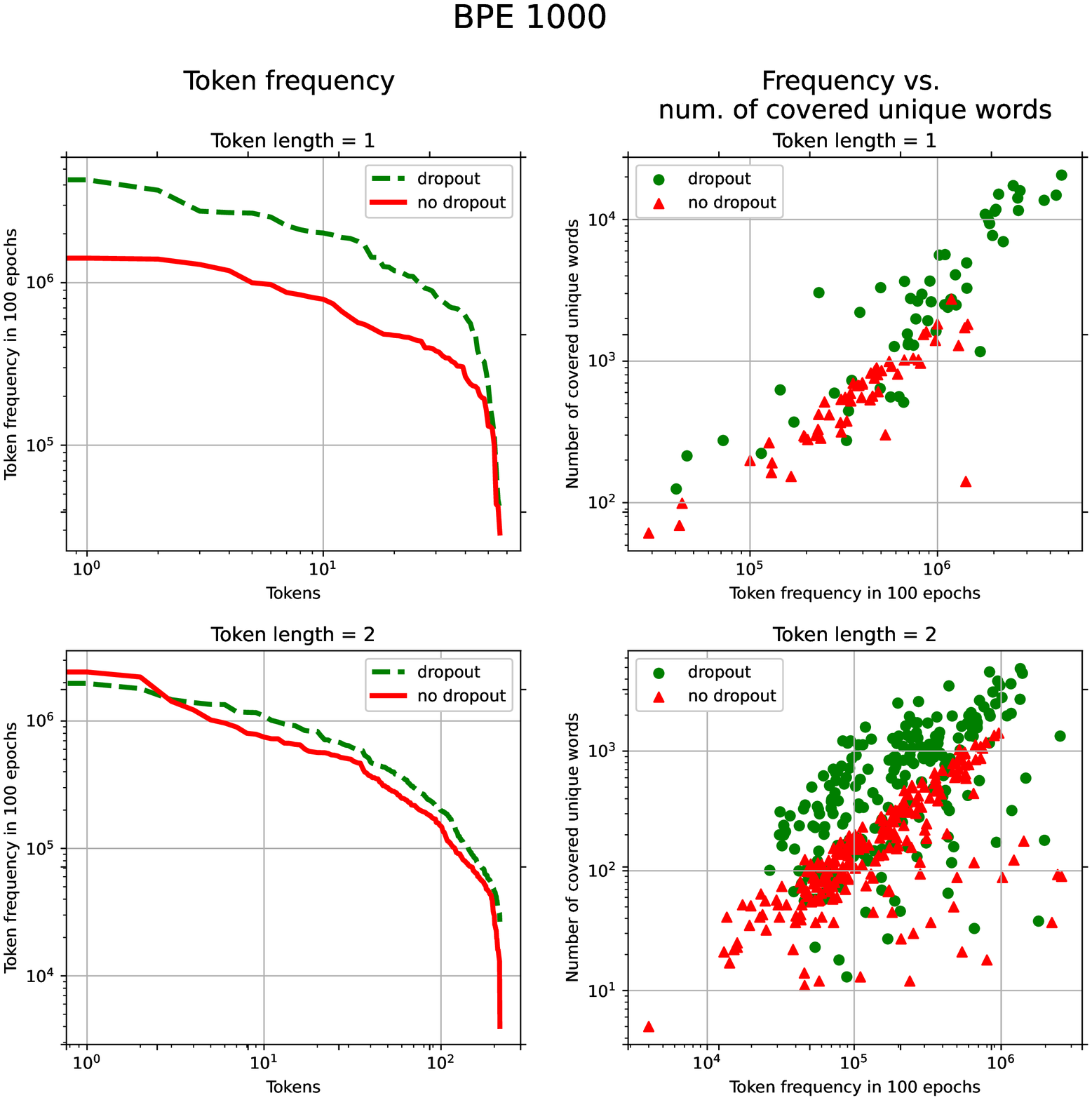}
  \caption{(Turkish): Left: Token frequency distribution in 100 epochs. The horizontal axis represents tokens sorted by their frequencies in the descending order. Vertical axis shows frequencies of tokens. Right: token frequency vs. the number of unique words in which these tokens are present. Points represent individual tokens. Both statistics were computed on the training set for token lengths 1 and 2 with the dropout and without. The BPE vocabulary size was set to 1000.}
  \label{fig:1001}
\end{figure*}

\begin{figure*}[ht]
  \centering
  \includegraphics[width=1\textwidth]{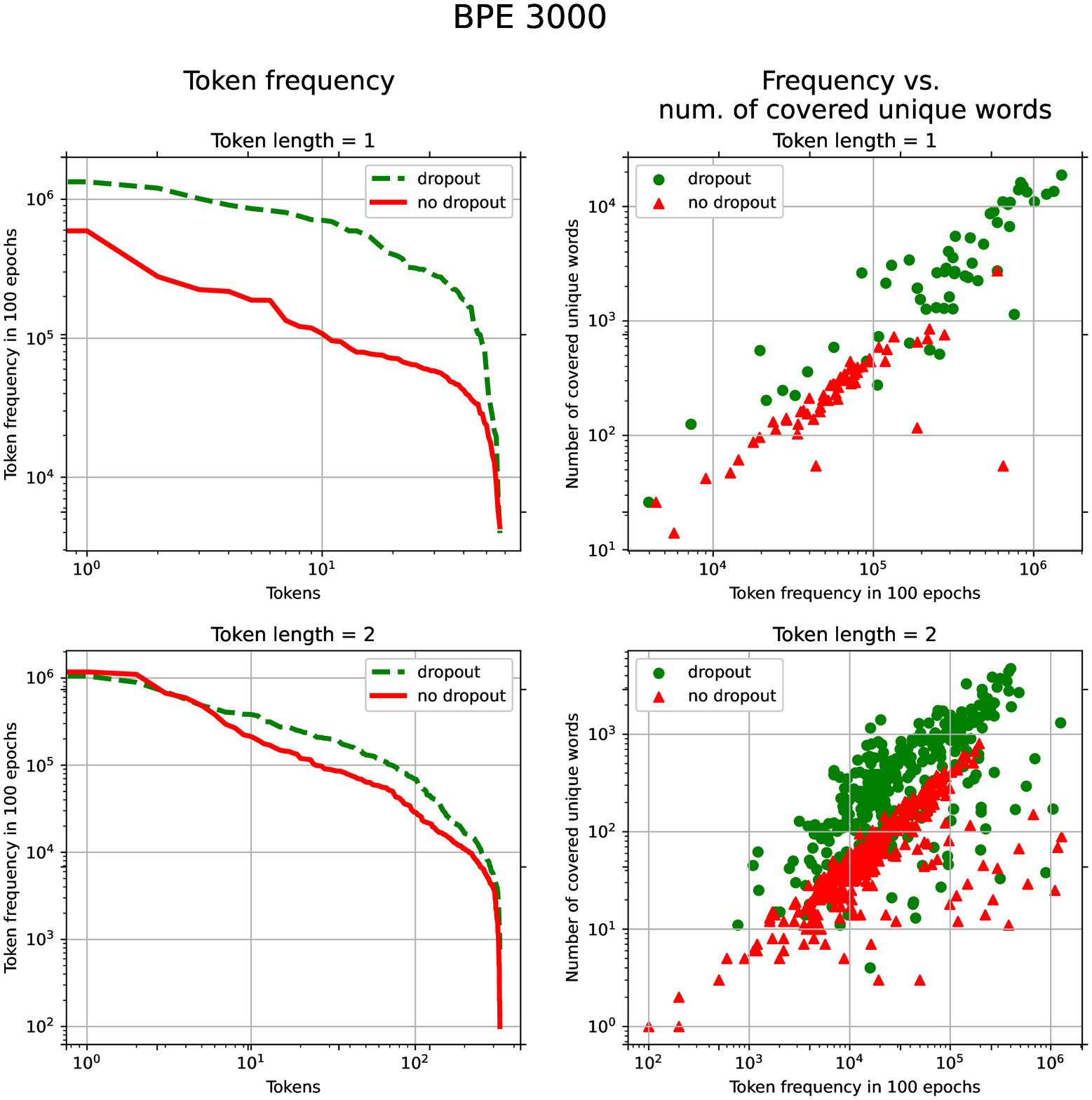}
  \caption{(Turkish): Left: Token frequency distribution in 100 epochs. The horizontal axis represents tokens sorted by their frequencies in the descending order. Vertical axis shows frequencies of tokens. Right: token frequency vs. the number of unique words in which these tokens are present. Points represent individual tokens. Both statistics were computed on the training set for token lengths 1 and 2 with the dropout and without. The BPE vocabulary size was set to 3000.}
  \label{fig:1002}
\end{figure*}

The increase in the frequency of short tokens occurs due to the ``forgetting'' to apply some merging rules when assembling short tokens into more complex ones. Without the ``forgetting'', these short non-terminal tokens would become a part of the other tokens. This leads to the situation where non-terminal tokens start appearing in words that were otherwise occupied by more advanced terminal subwords. Thus, the model receives more diverse contexts for these tokens during the training process. It can be seen in \Cref {fig:1001,fig:1002} that with BPE-dropout, short tokens were appearing evenly more often during the training (left line charts) and in a broader set of unique words (right scatter charts). The latter is also true for longer tokens (3-4 characters long).

We argue that short (1-2 characters long) tokens play an essential role in the recognition of OOV words. It was observed that their amount in the OOV recognition results ranges from 60-70\% to 80\%. Consequently, extensive statistics for short tokens and the variability of their contexts in training may help the model better produce unseen words based on the short token utterances already encountered in training.

Properties of BPE-dropout can be studied in terms of augmentation and regularization. For tokens 1-2 characters long, the method has strong augmentation properties. The use of BPE-dropout increased the amount of single-character tokens in the training process by 2-3 times: from 14 \% to 29 \% for the BPE vocabulary size of 1000 and from 7 \% to 22 \% for the vocabulary size of 3000. In other cases, BPE-dropout performed more like a regularization technique: the number of tokens with more than two symbols did not increase or even slightly decreased for tokens longer than four characters. For such subwords, diversification of token sequences is one of the regularization properties, as it reduces overfitting of the attention decoder.

\begin{figure*}[ht]
  \centering
  \includegraphics[scale=1]{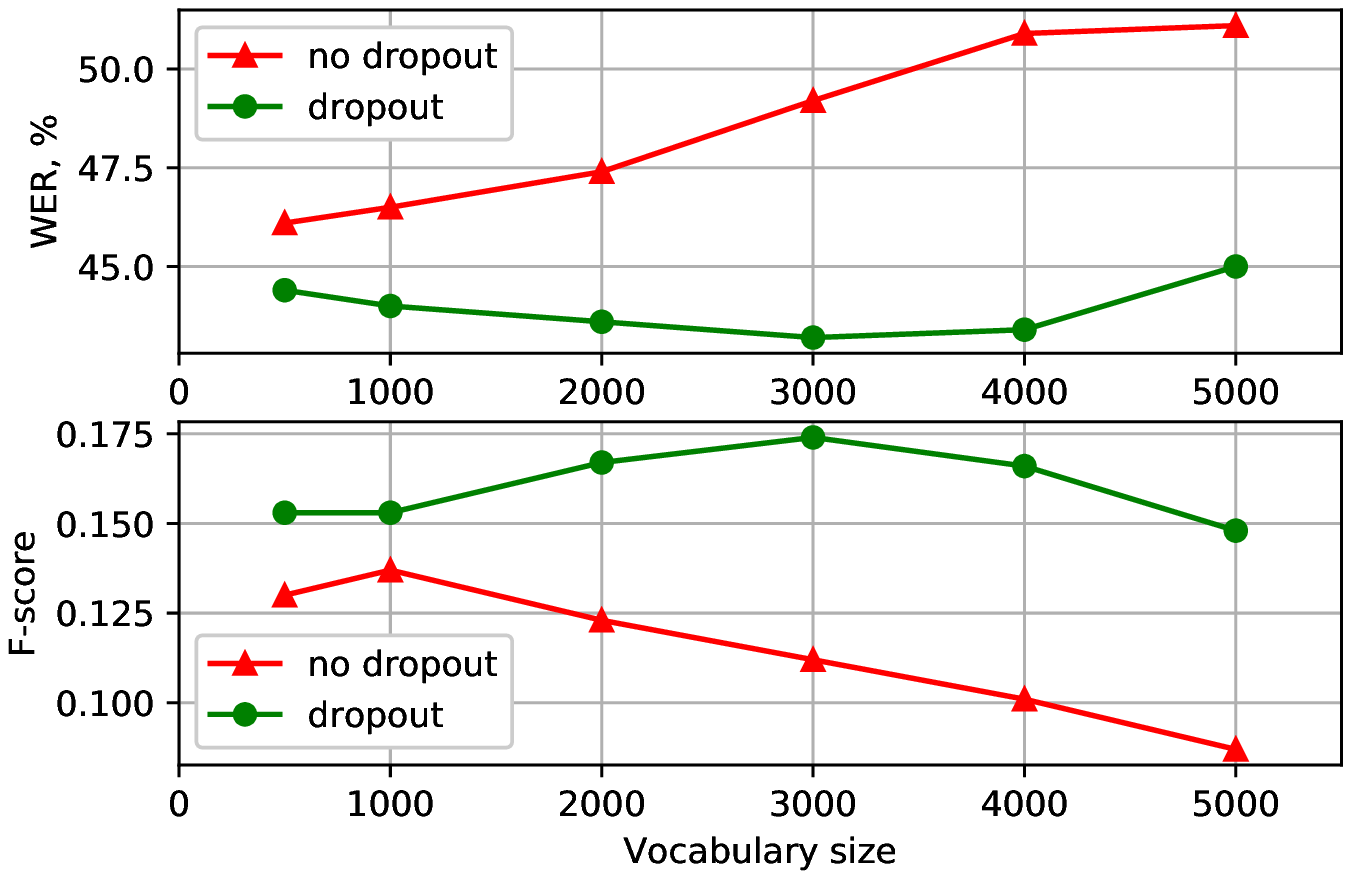}
  \caption{(Turkish): WER and F-score for different vocabulary size for BPE segmentation with the dropout (top) and without (bottom).}
  \label{fig:2}
\end{figure*}

\begin{figure*}[h!]
  \centering
  \includegraphics[scale=1]{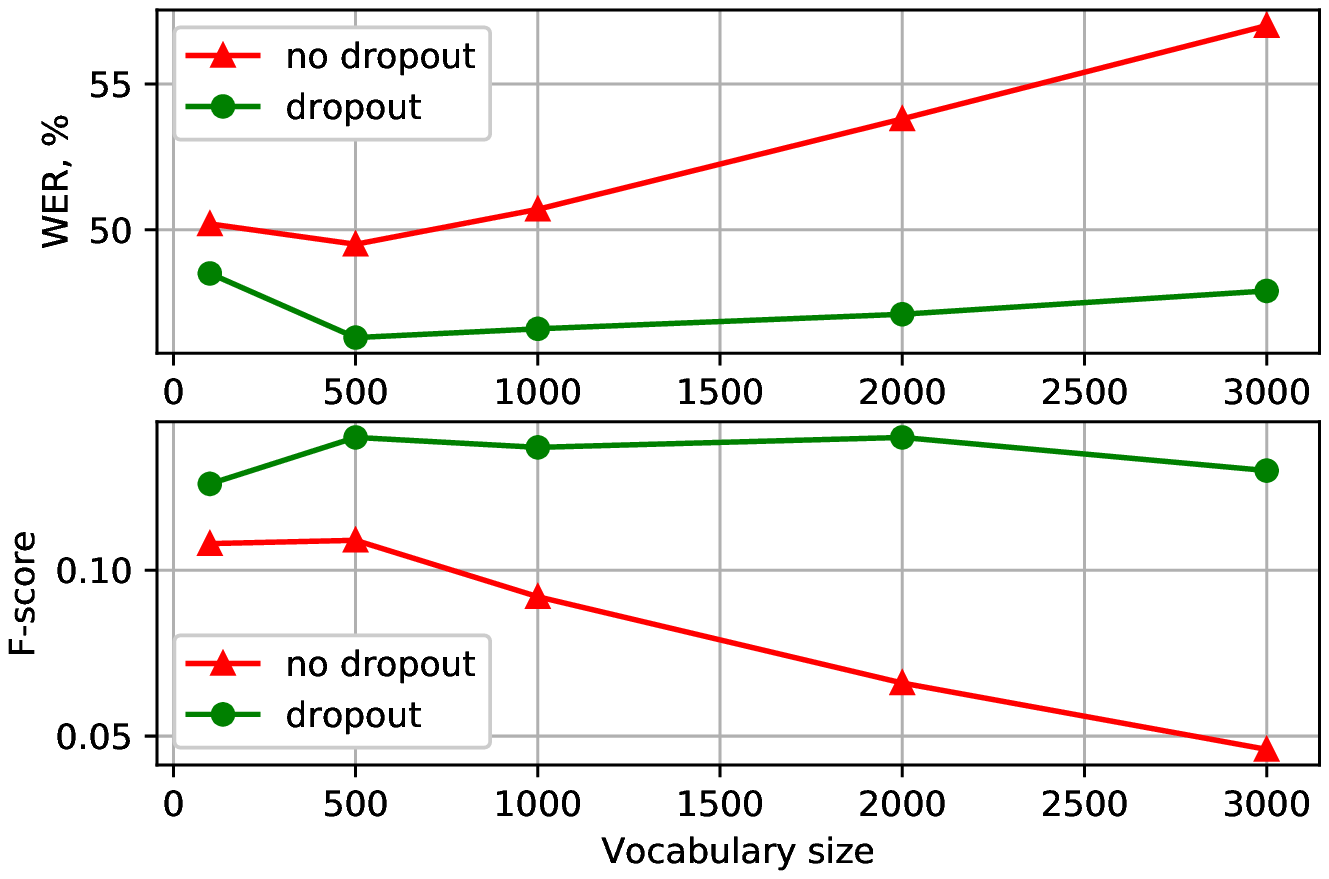}
  \caption{(Georgian): WER and F-score for different vocabulary size for BPE segmentation with the dropout (top) and without (bottom).}
  \label{fig:3}
\end{figure*}

\begin{figure*}[ht]
  \centering
  \includegraphics[width=1\textwidth]{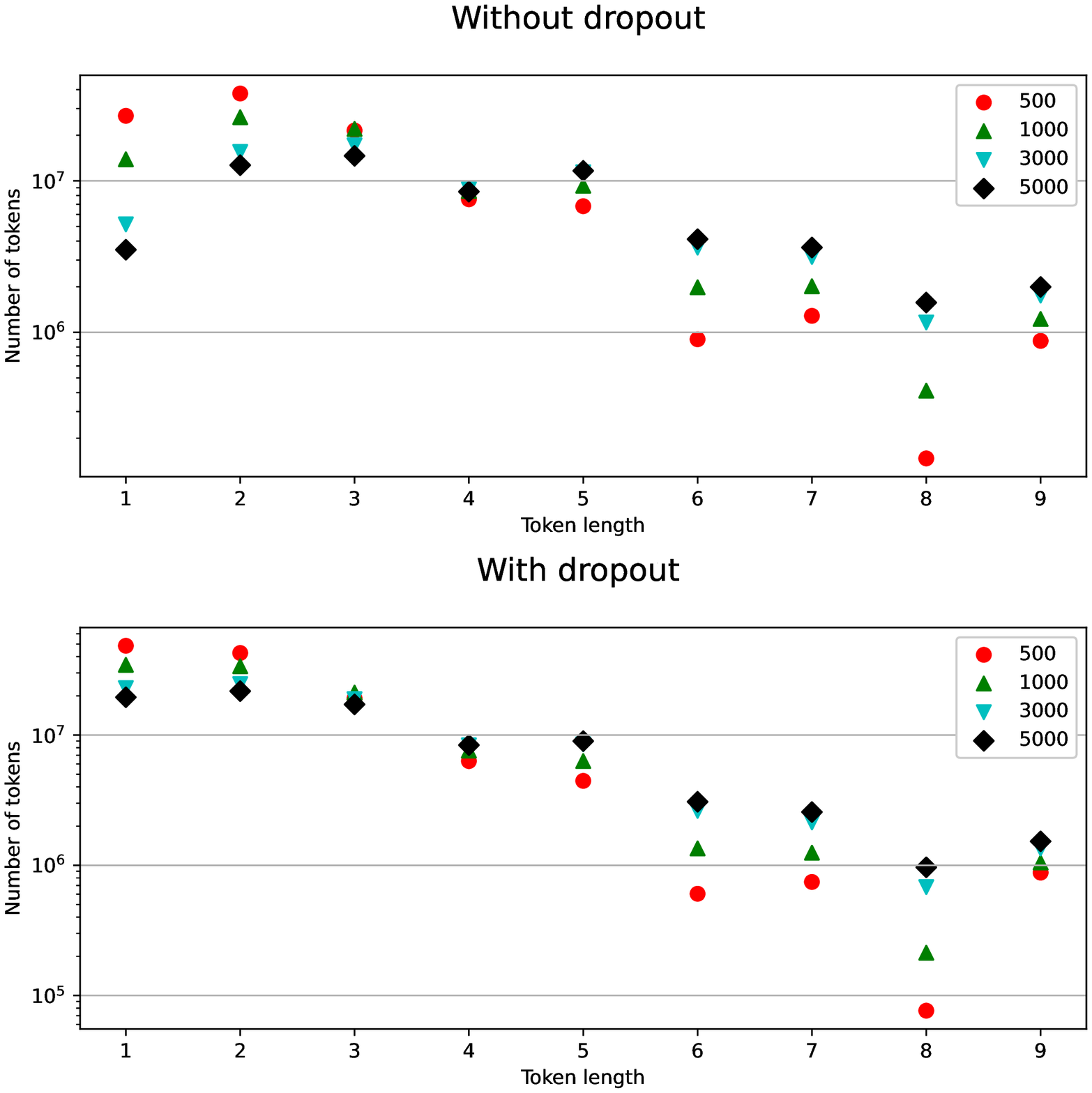}
  \caption{(Turkish): Number of tokens in 100 epochs vs. token length for different BPE subword vocabulary sizes with the dropout (bottom) and without (top).}
  \label{fig:1005}
\end{figure*}

Another important regularization property of BPE-dropout is a reduction of the influence of the vocabulary size on the model quality (according to \Cref{fig:2,fig:3}). A small vocabulary allows for better saturation of tokens when training, but the recognition may become non-robust to unconventional and alternative pronunciations, as modeling long-term language dependencies becomes difficult and acoustic information dominates decoding. Alternatively, increasing the BPE vocabulary allows more words to be recognized ``directly'' in one piece, which benefits the quality of recognition, but at the same time substantially shifts the balance of tokens in training towards long tokens, thereby obstructing the OOV recognition ability. The BPE-dropout technique facilitates the tradeoff between these options. As can be seen in \Cref{fig:1005}, BPE-dropout compensates for the decrease in the number of short token appearances at the cost of a slight decrease in the percentage of long ones.
 
\begin{figure*}[ht]
  \centering
  \includegraphics[width=1\textwidth]{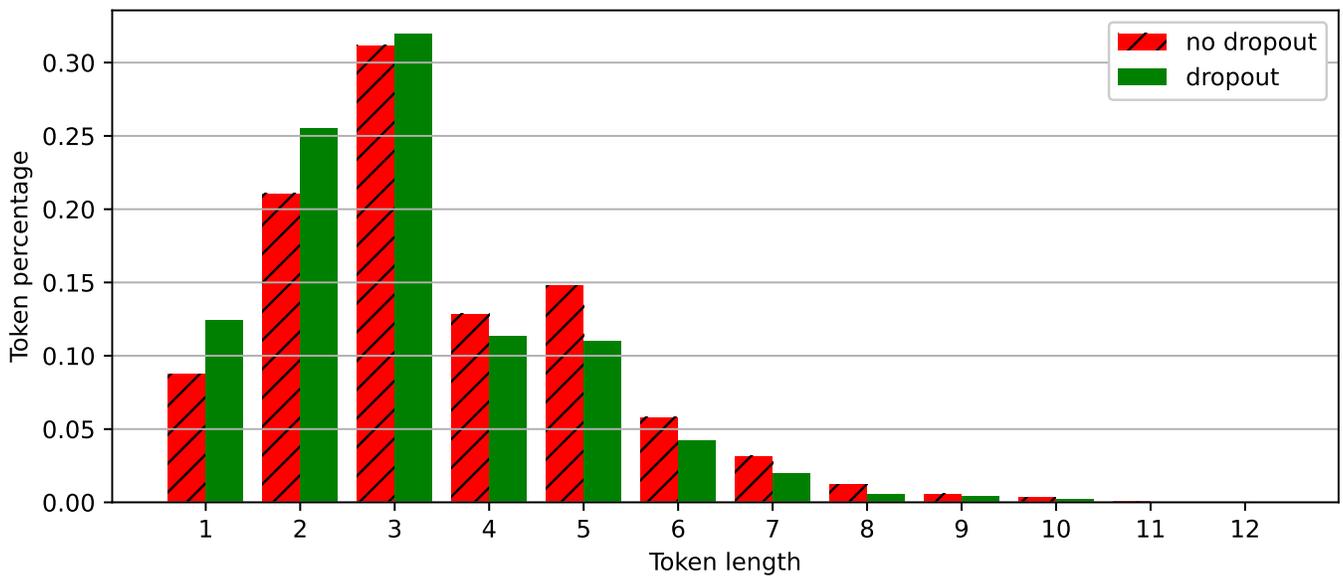}
  \caption{(Turkish): BPE token length distribution in OOV words emitted at decoding. Top: Vocabulary size 1000. Bottom: Vocabulary size 3000.}
  \label{fig:1003}
\end{figure*}

By revisiting \Cref{fig:2,fig:3}, it can be seen that the larger the vocabulary size, the more noticeable the improvement from using BPE-dropout augmentation. To explain this, we compared actual token distributions in recognition results. As demonstrated in \Cref{fig:1003}, BPE-dropout increases the number of relatively short (1-3 characters long) tokens in the OOV words from 61 \% to 75\%  for the BPE vocabulary size 3000. But in the case of the vocabulary size 1000, token length distributions in OOV are almost identical. This may mean that the improvement from the use of BPE-dropout is the greater, the more it reshapes and shifts the model token distribution towards shorter ones, assuming that the dropout parameter remains the same.

Overall, BPE-dropout-based augmentation provides the model with more complete and diverse statistics for tokens during the training, especially for the short ones. Besides, training with BPE-dropout allows the model to utilize a character-based model's properties to recognize OOV words while maintaining subword-based model quality for regular speech recognition tasks.

\section{Limitations}

Below are the main limitations of the study:
\begin{itemize}
    \item While the proposed acoustic unit augmentation approach significantly improves OOV recognition rate, it is still can not compete with or replace explicit personalization techniques for those personalized ASR tasks where quality is more important than speed.
    \item Since the use of BPE-dropout shifts the distribution of acoustic units towards shorter ones, the expected quality improvement might diminish if the method is applied to a system with a higher frame subsampling factor (e.g., 8 or 16).
    \item The end-to-end systems used in this study may not be suitable for use in smart assistants ``as is'', as the research focus was on quality improvement. Additional enhancements may be required to make them more efficient (e.g., model compression, decoding optimization, and streaming training mode).
    \item The data used in this study may not be sufficient to build an effective ASR system for smart assistants. It may require augmenting telephone waveforms with synthetic room impulse responses and extending it with target microphone data.
\end{itemize}

\section{Conclusion}
\label{sec:concl}

In this work, we proposed a method of dynamic acoustic unit augmentation based on the BPE-dropout technique. This method allows for improved ASR system quality at no additional training and decoding computational cost. Its regularization properties eliminate the need for optimal subword vocabulary size search, and its augmentation properties provide a consistent word error rate reduction (at least 6\% relative WER improvement compared to the best non-augmented models) in low-resource setups. Also, BPE-dropout's ability to significantly improve the recognition of out-of-vocabulary words makes it useful for personalized ASR tasks. Using this approach can make speech assistants' user experience better and improve the perceived quality. We found that our method is more effective than the previously used ULM subword regularization technique. Applying BPE-dropout unit augmentation to models trained on Babel Turkish and Georgian low-resource datasets helped our end-to-end monolingual models to be competitive with previous hybrid and multilingual systems.

Future work may concern adding Morfessor EM+Prune into consideration and comparison with BPE-dropout and the ULM subword regularization. Besides, non-deterministic subword tokenization should be studied in conjunction with the use of high frame subsampling factors. Finally, the dropout probability can be scheduled during the training to make a model behave differently (more like character- or purely subword-based) depending on the training stage.

\section*{Conflict of Interest Statement}

The authors declare that the research was conducted in the absence of any commercial or financial relationships that could be construed as a potential conflict of interest.

\section*{Author Contributions}
AL, AA, and IP contributed equally and share first authorship. IM and YM share senior authorship. AL developed the concept of the study. AL, AA, IP, and AM designed experiments. AA organized databases. AL and AA performed experiments. IP performed the post-experimental analysis. AL wrote the first draft of the manuscript, with significant contributions by IM and YM. AL, AA, and IP wrote sections of the manuscript. All authors contributed to manuscript revision, read, and approved the submitted version.

\section*{Funding}
This work was partially financially supported by ITMO University.

\section*{Data Availability Statement}
The data used in this study was obtained from Linguistic Data Consortium, Catalog \textnumero LDC2016S10\footnote{\url{https://catalog.ldc.upenn.edu/LDC2016S10}} and \textnumero LDC2016S12\footnote{\url{https://catalog.ldc.upenn.edu/LDC2016S12}}, the following restrictions apply\footnote{More in \url{https://catalog.ldc.upenn.edu/license/ldc-non-members-agreement.pdf}}: noncommercial linguistic education, research and technology development. Requests to access these datasets should be directed to Linguistic Data Consortium, \href{mailto:ldc@ldc.upenn.edu}{ldc@ldc.upenn.edu}.

\bibliography{mybib}

\section*{Figure captions}

\end{document}